# CRITICAL SIZES OF SATELLITE CONSTELLATIONS

Hugh G. Lewis

*Space Environment and Radio Engineering (SERENE) Group, School of Engineering, University of Birmingham, Edgbaston, Birmingham B15 2TT, United Kingdom, Email: h.g.lewis@bham.ac.uk*

**Abstract**

The precipitous growth of commercial space activity, largely from the development of satellite constellations, threatens to produce a runaway orbital debris population. National regulators are attempting to limit these risks through orbital debris mitigation requirements, but these fail to address the large scale and long lifetimes of satellite constellations. Instead, evaluations are based on the risks from single satellites, and the aggregate risks across constellations are neglected. This situation arises in part because of a lack of suitable approaches to evaluate constellation plans. Here a simple analytical approach reveals that several planned, partially deployed, or fully deployed constellations exceed their critical size producing an unstable or a runaway orbital debris population. Worryingly, this occurs both for small and large constellations, despite debris mitigation measures that go beyond internationally recognised good practices, and even when the constellations are deployed into a pristine, debris-free orbital environment. The results demonstrate the need for more stringent approaches to regulation and suggest careful consideration of some recent constellation plans.

## 1 Introduction

In the 5 years to 1st July, 2026, the number of active satellites in low earth orbit increased by nearly 12,000 as part of industry efforts to develop large satellite constellations[1]. These constellations currently offer imaging and communications services, including connections to the Internet for previously under-served populations. Plans for new constellations, surpassing 1.8 million satellites, seek to expand the services by providing data processing and renewable energy, amongst others[2]. Constellations share some characteristics of terrestrial infrastructure systems, including their long lifetimes, large scale and cost, influence on the wider economy, and extensive impacts on society and the environment[3]. As with terrestrial infrastructure systems, the environmental impacts of orbital infrastructure systems such as satellite constellations need to be addressed through appropriate standards and regulations. However, constellation proposals and deployments are proceeding at a pace outstripping that of standards development at the potential cost of the environment. In 2026 the Federal Communications Commission (FCC) has already received applications for new constellations comprising more than 1.2 million satellites[4], which would increase the existing number of operational satellites by more than 7800%[1].

The stability model by Kessler and Anz-Meador identified orbital regions near 900 km and 1400 km where the number of intact objects in the February 1999 satellite catalogue already exceeded the critical number needed for a runaway environment, as characterised by a collision rate sufficient to drive the population of fragments to infinity over an infinite amount of time[5]. A revised stability model evaluated the March 2025 satellite catalogue, revealing that the population of intact objects exceeded the critical number across an expanded range of altitudes between 520 km and 1000 km[6]. This change was likely driven by the development of satellite constellations in the period between the two studies[1]. However, a key assumption made by both stability models was that orbital objects are incapable of manoeuvring to mitigate collision risk. Importantly, the Starlink constellation made up a substantial proportion of the March 2025 catalogue, and yet component satellites were demonstrably capable of collision risk mitigation manoeuvres[7]. This is likely to have created an overly pessimistic outlook in the 2025 study.

Consequently, an appropriate methodology is urgently needed to evaluate constellation plans and to support standards development and regulation. The stability model developed by Kessler and Anz-Meador[5], and subsequently updated by Lewis and Kessler[6], provides the basis of a simple analytical solution, by defining the environment at a point in space and calculating only the rate that objects pass through that point. In this new work, the original stability model is adjusted, accounting for constellation operational practices including collision risk mitigation and post-mission disposal[8], to calculate the critical sizes of various constellations. Each constellation is considered in isolation, neglecting existing satellite and debris populations and other constellation plans. This offers an optimistic view. In this context, a fundamental principle should be that the constellations remain below these critical sizes and do not cause an unstable or runaway environment.

## 2 Results

The characteristics of 14 constellations that are planned, partially deployed, or are already fully deployed in low earth orbit (Tab. 1), comprising nearly 1.7 million satellites in total, are evaluated using the new stability model to determine their critical sizes.

**Tab. 1**. Planned and deployed constellation sizes and satellite characteristics used in this study. Constellations are ordered largest to smallest. The rightmost column identifies the satellites used to parameterise the stability model (see "Methods").

| Constellation | Number of Satellites | Satellite Mass (kg) | Satellite Area ($m^2$) | Satellite A/m ($m^2$/kg) | Satellite (NORAD ID) |
|---|---|---|---|---|---|
| Starmind (SpaceX Orbital Data Centers)[9] | 1,000,000 | 575 | 36.964 | 0.0643 | Starlink 37240 (69031) |
| E-Space Cinnamon & Semaphore[2] | 453,963 | 10 | 0.126 | 0.0126 | E-Space 1 (52423) |
| Starlink Generation 3[10] | 100,000 | 575 | 36.964 | 0.0643 | Starlink 37240 (69031) |
| Starcloud[11] | 87,996 | 50 | 0.225 | 0.0045 | Starcloud-1 (66303) |
| Starlink Mobile Satellite Service[12] | 15,000 | 575 | 36.964 | 0.0643 | Starlink 37240 (69031) |
| Guowang[2] | 12,992 | 800 | 5.935 | 0.0074 | Hulianwang Weixing Digui 16-03 (67061) |
| Amazon Leo[13] | 7,774 | 490 | 2.326 | 0.0048 | Kuiper 00261 (67158) |
| Blue Origin TeraWave[14] | 5,280 | 490 | 2.326 | 0.0048 | Kuiper 00261 (67158) |
| Amazon Leo Direct-to-Device System[15] | 5,105 | 490 | 2.326 | 0.0048 | Kuiper 00261 (67158) |
| Lynk[2,16] | 2,000 | 85 | 0.65 | 0.0076 | Lynk Tower 8 (69014) |
| Eutelsat OneWeb[17,18] | 637 | 148 | 3.464 | 0.0234 | OneWeb SL0706 (61611) |
| Eutelsat Next[19] | 528 | 148 | 3.464 | 0.0234 | OneWeb SL0706 (61611) |
| Telesat Lightspeed[2] | 300 | 30 | 0.28 | 0.0093 | Telesat LEO 3 (57392) |
| AST SpaceMobile[20] | 243 | 5850 | 24.992 | 0.0043 | SpaceMobile 006 (67232) |

Fig. 1 compares these constellations. The largest constellation considered is the fleet of SpaceX orbital data centres called 'Starmind' (1 million satellites planned)[9] and the smallest is the fleet of AST SpaceMobile satellites (243 satellites planned)[20]. The largest four constellations represent 97% of the total number of satellites. The average constellation size is 120,843 satellites. The median constellation size is 6,527 satellites.

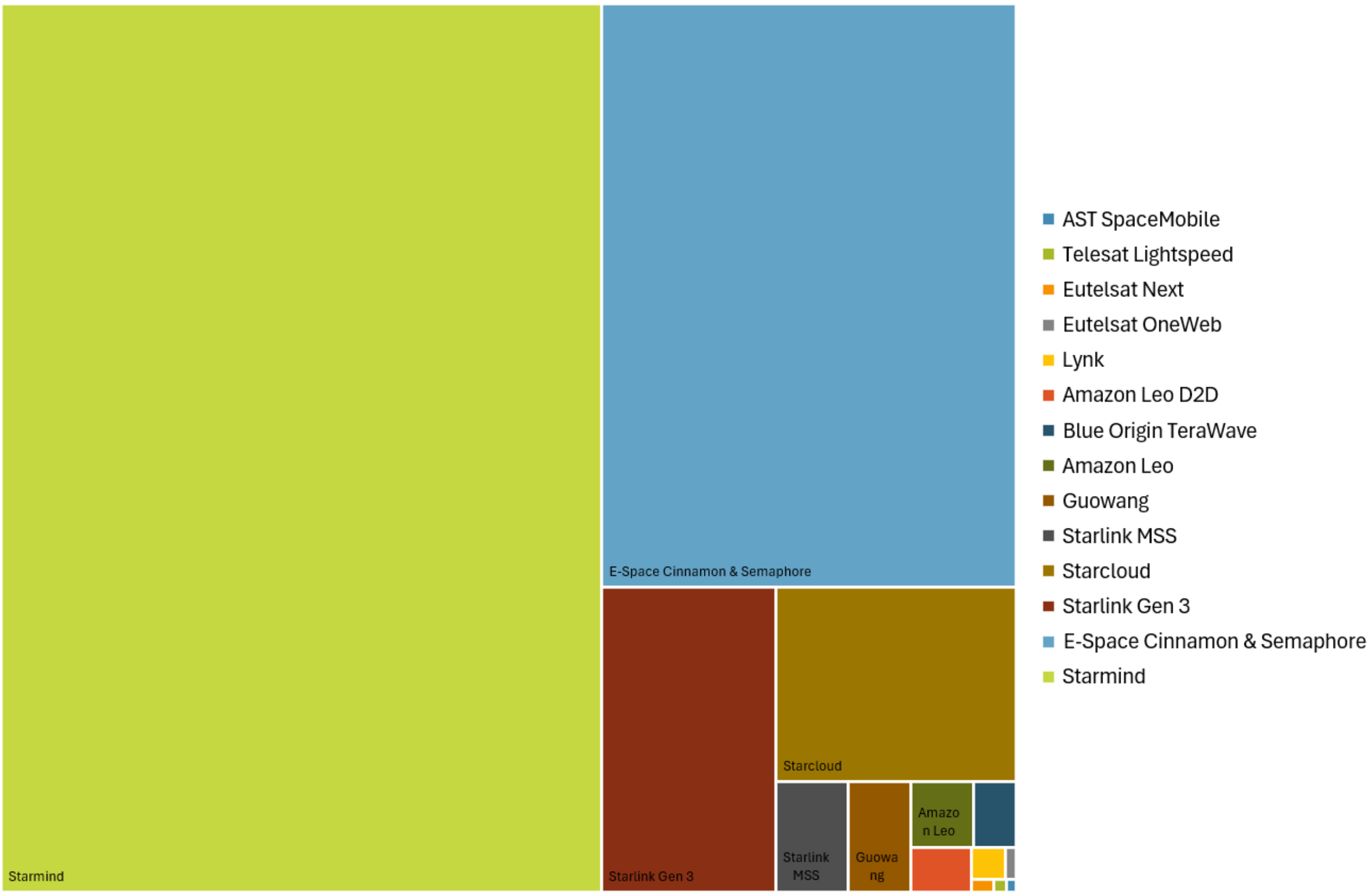

**Fig. 1**. Comparison of the sizes of 14 constellations that are planned, partially deployed, or fully deployed in low earth orbit. The area of each element shown in the figure is directly proportional to the number of satellites in the respective constellation.

Plots of the critical constellation sizes for runaway and unstable fragment populations as functions of altitude are shown in Figs. 2-4 for the constellations (see "Methods"). The red lines in Figs. 2-4 represent the critical sizes of the constellations leading to a runaway environment. These lines are the maximum number of satellites that can be maintained in the constellation before a runaway fragment population is generated. The amber lines represent the critical sizes leading to an unstable environment. The solid lines assume a "control-to-re-entry"[21] post-mission disposal option and the dashed lines assume the FCC's "5-year rule" post-mission disposal option[22].

Fig. 2 shows constellations that exceed the critical size for a runaway environment. Fig. 3 shows constellations that exceed the critical size for an unstable environment. Fig. 4 shows constellations that are below the critical sizes for unstable and runaway environments. If a constellation is above the runaway threshold for the given altitude, the breakup of satellites is predicted to produce a fragment population that will tend to grow without limit over an infinite amount of time even without other objects in the environment[5,6]. Constellations above the unstable threshold for the given altitude are predicted to produce a fragment population that will tend to grow to some equilibrium size[5,6]. Constellations below both thresholds might still experience satellite breakups, but the fragment population will not tend to increase.

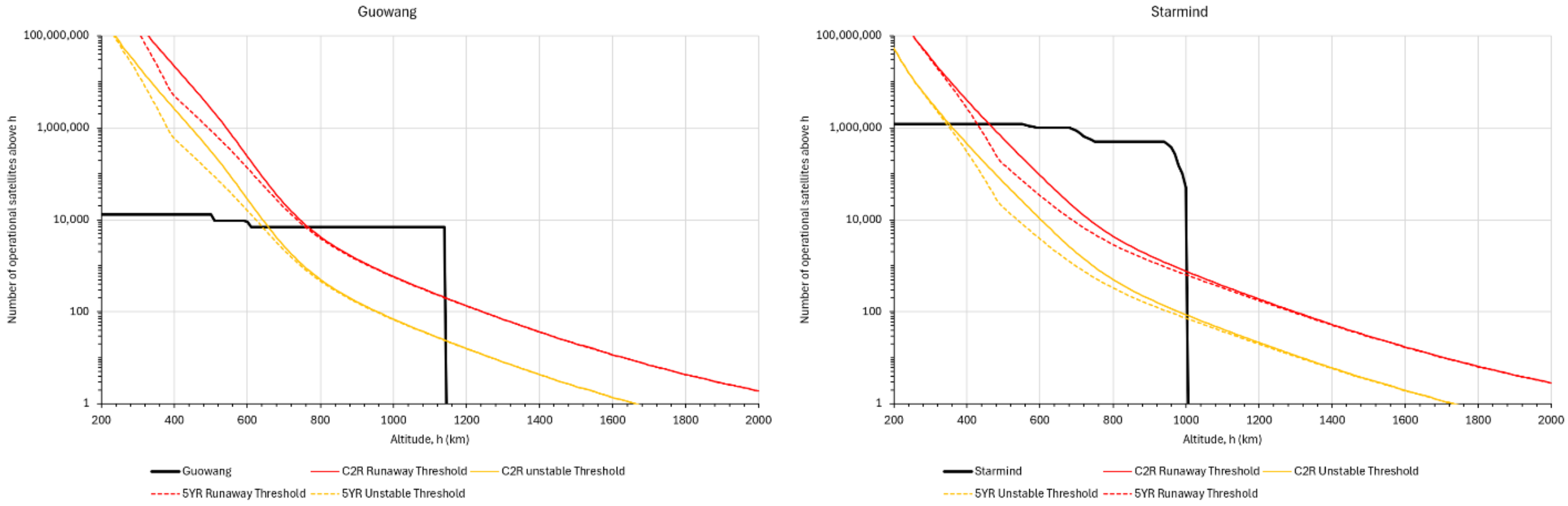

**Fig. 2**. Two constellations predicted to exceed their respective unstable (amber) and runaway (red) thresholds, meaning the fragment populations are predicted to grow to an infinite size over an infinite time, assuming either control-to-re-entry (solid lines) or 5-year rule (dashed lines) post-mission disposal options.

Telesat Lightspeed
Eutelsat Next
Eutelsat OneWeb
Starcloud
Starlink Gen 3
Number of operational satellites above h
Altitude, h (km)
Altitude (km)
Telesat
Eutelsat Next
Eutelsat OneWeb
Starcloud
Starlink Gen3
C2R Runaway Threshold
C2R Unstable Threshold
5YR Runaway Threshold
5YR Unstable Threshold

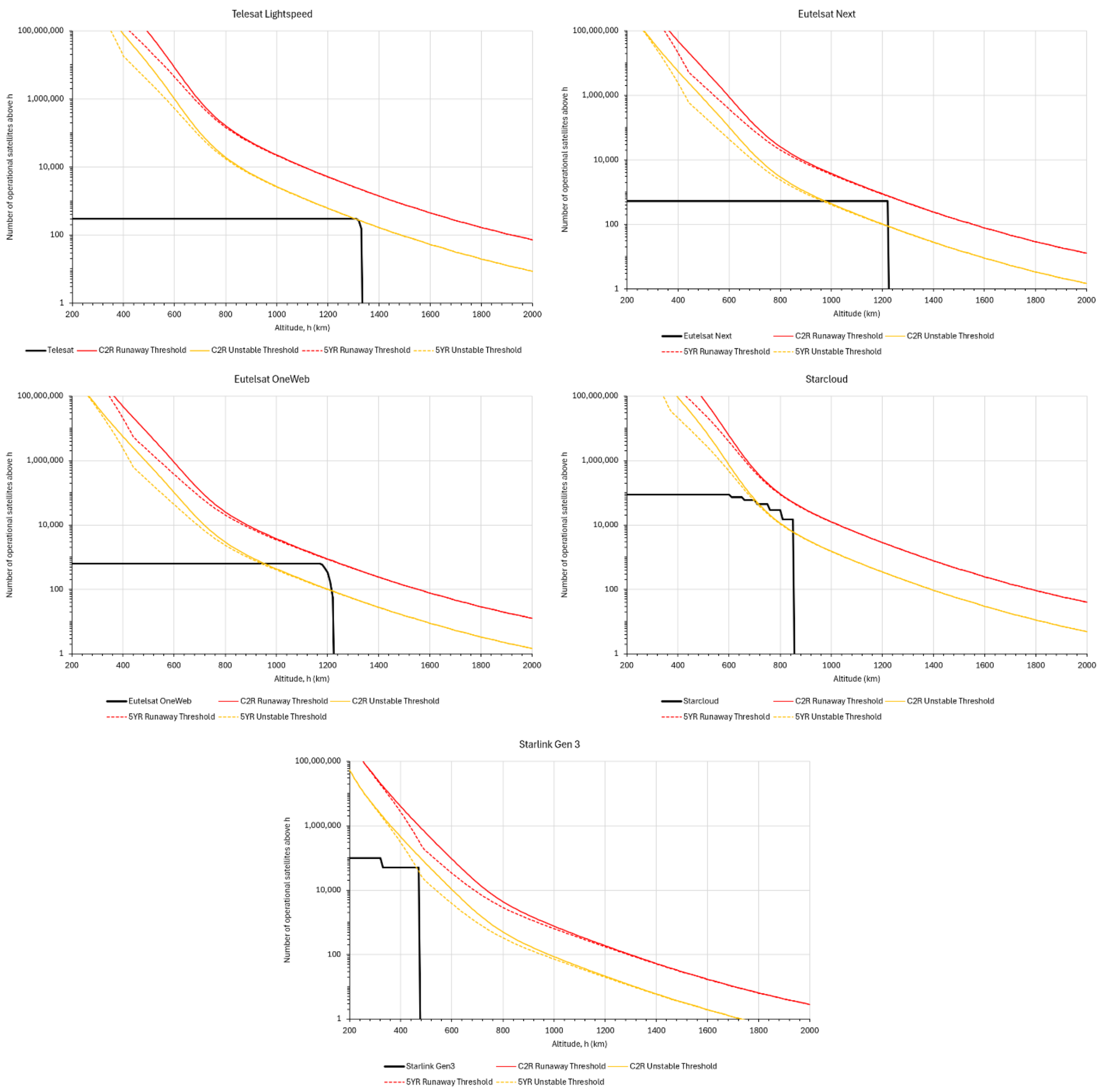

**Fig. 3**. Five constellations predicted to fall below their respective runaway thresholds (red) but exceed their unstable thresholds (amber), meaning the fragment populations are predicted to grow to an equilibrium level, assuming either control-to-re-entry (solid lines) or 5-year rule (dashed lines) post-mission disposal options.

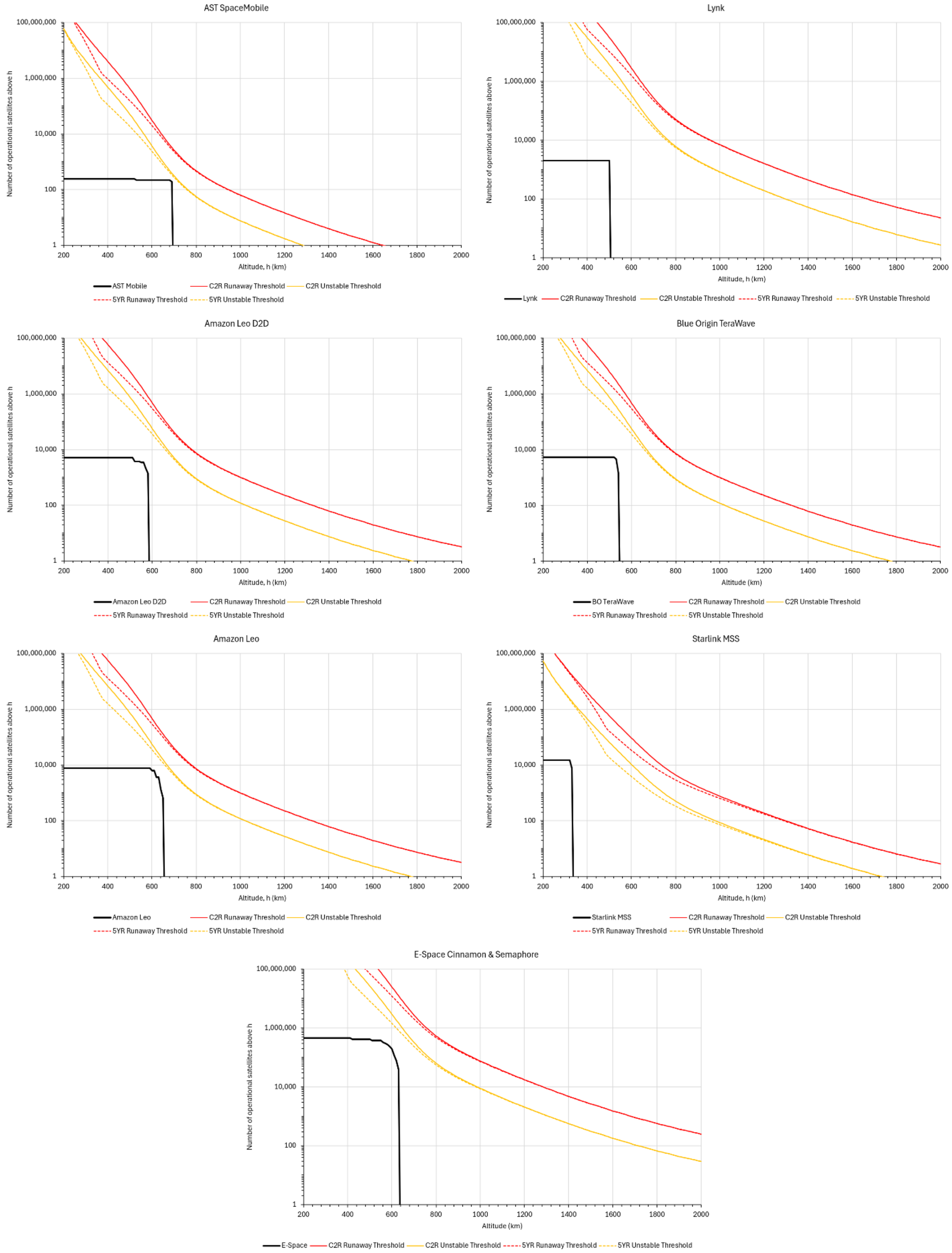

**Fig. 4**. Seven constellations predicted to fall below their respective unstable (amber) and runaway (red) thresholds, meaning the fragment populations are predicted to decline, assuming either control-to-re-entry (solid lines) or 5-year rule (dashed lines) post-mission disposal options.

Starcloud just exceeds the "control-to-re-entry" runaway threshold if the satellite mass and area are doubled with respect to the values for Starcloud-1 (object 66303). If Starcloud deploys satellites with the same mass and area as Starlink 37240 (69031) or the substantially larger Starship version[23], the runaway and unstable thresholds are exceeded (Fig. 5). Similarly, Eutelsat Next is close to its critical size for runaway assuming satellite characteristics equal to those of Eutelsat OneWeb. If the Eutelsat Next satellites have relatively small increases in mass and area, compared with the OneWeb satellites, the runaway threshold is also exceeded. Additionally, combining the Eutelsat OneWeb and Eutelsat Next constellations results in the merged system exceeding the critical size for a runaway environment even with an assumption of the smaller OneWeb satellite characteristics. However, the merged system falls below the runaway threshold if the satellite mission lifetimes are assumed to be 8 years instead of 5 years. The Telesat Lightspeed constellation also falls below the unstable and runaway thresholds if the mission lifetime is assumed to be 8 years instead of 5 years (Fig. 5). Similar combinations of the Amazon Leo and Amazon Leo D2D System and the Starlink Generation 3 and Starlink MSS do not change the outcomes for their owner/operators. The combined E-Space Cinnamon and Semaphore constellation exceeds the runaway threshold at altitudes above 360 km if the 10 kg nanosatellites are assumed to have no propulsion capability[24] (Fig. 5). Fewer than 100 of the nanosatellites can be maintained at the highest altitude proposed for the constellation.

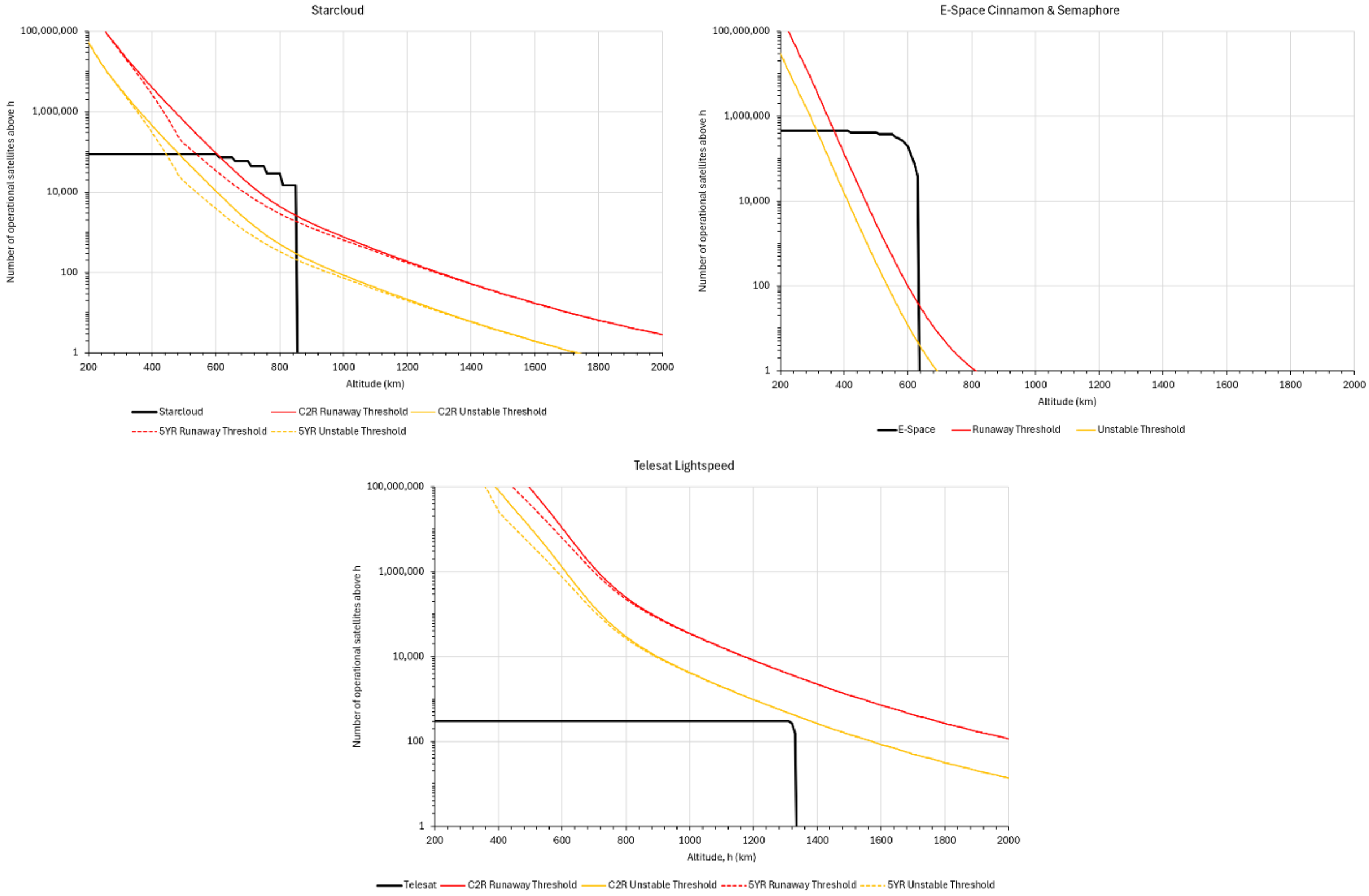

**Fig. 5**. Starcloud and the E-Space Cinnamon and Semaphore constellations are predicted to exceed their respective unstable (amber) and runaway (red) thresholds after accounting for appropriate satellite characteristics. There is no difference between the "control-to-re-entry" and "5-year rule" for the E-Space constellation for the case without propulsion. The Telesat Lightspeed constellation is predicted to fall below its unstable and runaway thresholds if the mission lifetime is 8 years rather than 5 years.

The application by Blue Origin to the FCC for its Project Sunrise constellation, which comprises 51,600 satellites operating in orbits from 500-1,800 km, does not include sufficient detail about the orbital configuration or the satellites to be included in the assessment above[25]. However, the application states, "Each orbital plane will contain approximately 300-1,000 satellites," so it is possible to apply the stability model after making some assumptions. Firstly, assume the satellites are equal in mass and area to the Kuiper 00261 satellite (67158) adopted for the Amazon Leo, Amazon Leo D2D System, and Blue Origin TeraWave (Tab. 1, and "Methods"). Secondly, assume that 300 of these satellites are deployed in a discrete altitude band at 1,800 km altitude. Application of the stability model for these assumptions shows the unstable threshold is 7.3 satellites and the runaway threshold is less than 1 satellite at 1,800 km, hence the Project Sunrise constellation will likely exceed the critical size even for satellites with substantially different mass, area, and mission lifetime.

Similarly, the application to the FCC by Cowboy Space Corporation specifies that its Stampede constellation, "will consist of up to 20,000 satellites operating in dawn dusk sun-synchronous orbits ("SSO") and contained in orbital shells between 700 and 1000 km." Adopting satellite characteristics equal to those of Kuiper 00261 (67158), the critical number of satellites for a runaway environment at 700 km is 40,272 and the critical number for an unstable environment is 4,845. Hence, the Stampede constellation would be above its critical size for an unstable environment. However, if the satellite characteristics of Starlink 37240 (69031) are used, the critical number for a runaway environment becomes 16,824 and the Stampede constellation would exceed this.

After accounting for different satellite characteristics and these additional constellations, the results show six constellations may exceed critical sizes for a runaway environment: Guowang, Stampede, Project Sunrise, Starcloud, the combined E-Space Cinnamon and Semaphore, and Starmind. Under the model assumptions, these constellations are predicted to produce an ever-increasing fragment population despite adherence to regulations, the adoption of debris mitigation measures that go beyond internationally recognised good practices and the use of effective risk mitigation manoeuvres[8]. Three constellations are predicted to exceed critical sizes for an unstable environment: Eutelsat OneWeb, Eutelsat Next, and Starlink Generation 3. Seven constellations are predicted to be below the critical sizes for unstable and runaway environments: AST SpaceMobile, Telesat Lightspeed, Lynk, Amazon Leo Direct-to-Device (D2D) System, Blue Origin TeraWave, Amazon Leo, and Starlink Mobile Satellite Service (MSS).

## 3 Discussion

National regulators typically require orbital debris mitigation plans for new satellite systems[27], but these rarely address aggregate risks arising from the planned deployment of many satellites into constellations. Instead, they focus on the risks associated with a single, non-manoeuvrable satellite. The risk of collisions with large objects is considered to be zero while satellites are manoeuvrable[27]. Additionally, regulators seldom recognise that constellations have characteristics of orbital infrastructure systems, which require recurring replenishment and disposal of large numbers of satellites to achieve long service lifetimes. This means that substantial populations of satellites are ever-present in the orbital environment but are not counted among the user-serving constellation. If the orbital debris mitigation plan is consistent with either the IADC's "25-year rule"[28] or the FCC's "5-year rule"[29] then a large proportion of these additional satellite populations will have no capability to perform collision risk mitigation manoeuvres and may pose a high collision risk.

Constellation operators/owners rightly argue that they are meeting all requirements for orbital debris mitigation set out by their national regulators, and some could say they are going beyond what is required[30]. However, there are gaps in our understanding of aggregate risks, both in terms of how to measure them, and how to apply them in regulation[31]. With recent applications for truly "mega-constellations"[9], setting a precedent for more to follow in the future, there is a danger that licences granted to these based on the seemingly negligible risks associated with a single satellite could place the evolution of the orbital environment onto a difficult trajectory.

With the introduction of a stability model, modified from the one presented by Lewis and Kessler[6], to capture constellation characteristics, this study has demonstrated an ability to evaluate constellation plans easily and appropriately. The approach needs no additional information from the owner/operator beyond what they already provide with their application, and it makes no assumptions about existing satellite or debris populations in the orbital environment. Each constellation is evaluated against its own critical size, with a binary outcome (i.e., "pass" or "fail"). To demonstrate the approach, it was applied to 14 planned, partially deployed, or fully deployed constellations, parameterised using data from the FCC and the European Space Agency's (ESA's) Database and Information System Characterising Objects in Space (DISCOS)[32] and data curated by Jonathan McDowell[2]. One other constellation, Blue Origin Project Sunrise[25], was evaluated based on a fraction of its satellites in its highest identified altitude band. Additionally, Cowboy Space's Stampede constellation[26] was evaluated based on the number of satellites above its minimum specified altitude.

The results show that nine of these 16 constellations exceed their respective critical sizes, after reasonable modifications to their satellite characteristics. They are predicted to generate fragment populations that grow either without limit, or potentially to a high equilibrium level, even without contributions from other orbital objects or the other constellations. This means that more than half of the constellations investigated may individually pose a threat to safe and sustainable satellite operations in the orbital environment, even after accounting for good operating practices. The orbital debris mitigation plans for constellations falling into this category and still under review by national regulators, should be evaluated rigorously and their approval seriously considered.

The diversity of the constellations approaching or exceeding the critical size suggest an evaluation against orbital debris criteria cannot be based on the number of satellites alone, or on the physical characteristics of or risk associated with a single satellite. For instance, the Eutelsat Next constellation[19], with 528 satellites, is among the smallest constellations investigated and assumed to use relatively small satellites, yet it exceeds its unstable threshold. By comparison, Starlink MSS[12] with 15,000 satellites, is larger than twice the median constellation size and uses substantially larger satellites but is below its unstable and runaway thresholds. The constellation size and orbital configuration, and satellite physical characteristics and manoeuvre capability combine in the stability model to inform the runaway and unstable threshold limits.

The outcomes of the evaluations above may be sensitive to some assumptions made regarding these parameters. However, when the constellations are far above or below the unstable or runaway thresholds, even relatively large changes to these parameters do not affect the outcome of the assessment using the stability model. For example, SpaceX has stated[9] the mass of a Starmind satellite is 3,000 kg and has a total area of 800 $m^2$, but the evaluation above uses the values from DISCOS for a substantially smaller Starlink satellite. A repeat evaluation based on the larger satellite does not change the outcome – the constellation remains above its critical size for a runaway environment. The same is true for the Blue Origin Project Sunrise and the Cowboy Space Stampede constellations after assuming smaller satellites or longer mission lifetimes. Yet, the removal of manoeuvre capability among satellites of the E-Space Cinnamon and Semaphore constellations did nonetheless generate substantially different thresholds and reverse the original positive outcome. Generally, changes to satellite physical characteristics and capabilities can lead to different outcomes when the constellations are near the unstable or runaway thresholds, as shown for Starcloud, Eutelsat OneWeb, Eutelsat Next, and Telesat Lightspeed. Owners/operators are not reliant upon assumptions and have the advantage of direct access to and control of these parameters, meaning they can select orbital configurations and satellite designs to demonstrate desirable outcomes to regulators. Regulators can easily verify such outcomes or offer guidance on which parameters ought to be adjusted.

The stability model produces results that can be optimistic. Firstly, each constellation is evaluated separately and under the assumption that no other orbital object is present in the environment. In this context, a constellation exceeding its critical size will see breakups only of its own satellites and the self-generation of a growing population of fragments. If other satellites and debris objects – including rocket stages used to deploy the constellation, which may be abandoned in orbit[33] – were included, the critical constellation sizes would be lower than shown and more of the constellations would exceed these. Although the model neglects other satellites and debris in the orbital environment, and the cumulative effects arising from deployments of multiple constellations, these influences should not be forgotten in any appraisal for licensing purposes. The consequence of ignoring them is an increased potential for an unstable or runaway environment. Secondly, the model excludes non-collisional breakups and fragments below the minimum mass needed to catastrophically breakup an intact satellite, even though they could still cause a satellite breakup. Inclusion of these factors would also reduce the critical size of the constellations. Thirdly, the stability model assumes the constellation is distributed uniformly within a spherical shell (see Fig. 6 (a) and (b)). However, the satellites in all the constellations investigated are deployed into orbits with orbital inclinations that result in the satellites – and any fragments they generate – spending all their time within spherical shell segments (Fig. 6 (c), (d), and (e)). The volume of a spherical shell segment is proportional to the Sine of the orbital inclination, leading to a smaller volume than the assumed spherical shell (see "Methods") and reducing the critical size of the constellation if adopted in the stability model[1]. Constellations designed to provide orbital data centre services are constrained to Sun-synchronous orbits with specific Local Solar Time (LST) ascending nodes, which further reduces the volume occupied by the user-serving satellites.

[1] Doing so would add substantial complexity to the method, as it would essentially require a separate stability model for each shell of the constellation with a different orbital inclination, and a method for addressing the intersection of the spherical shell segments.

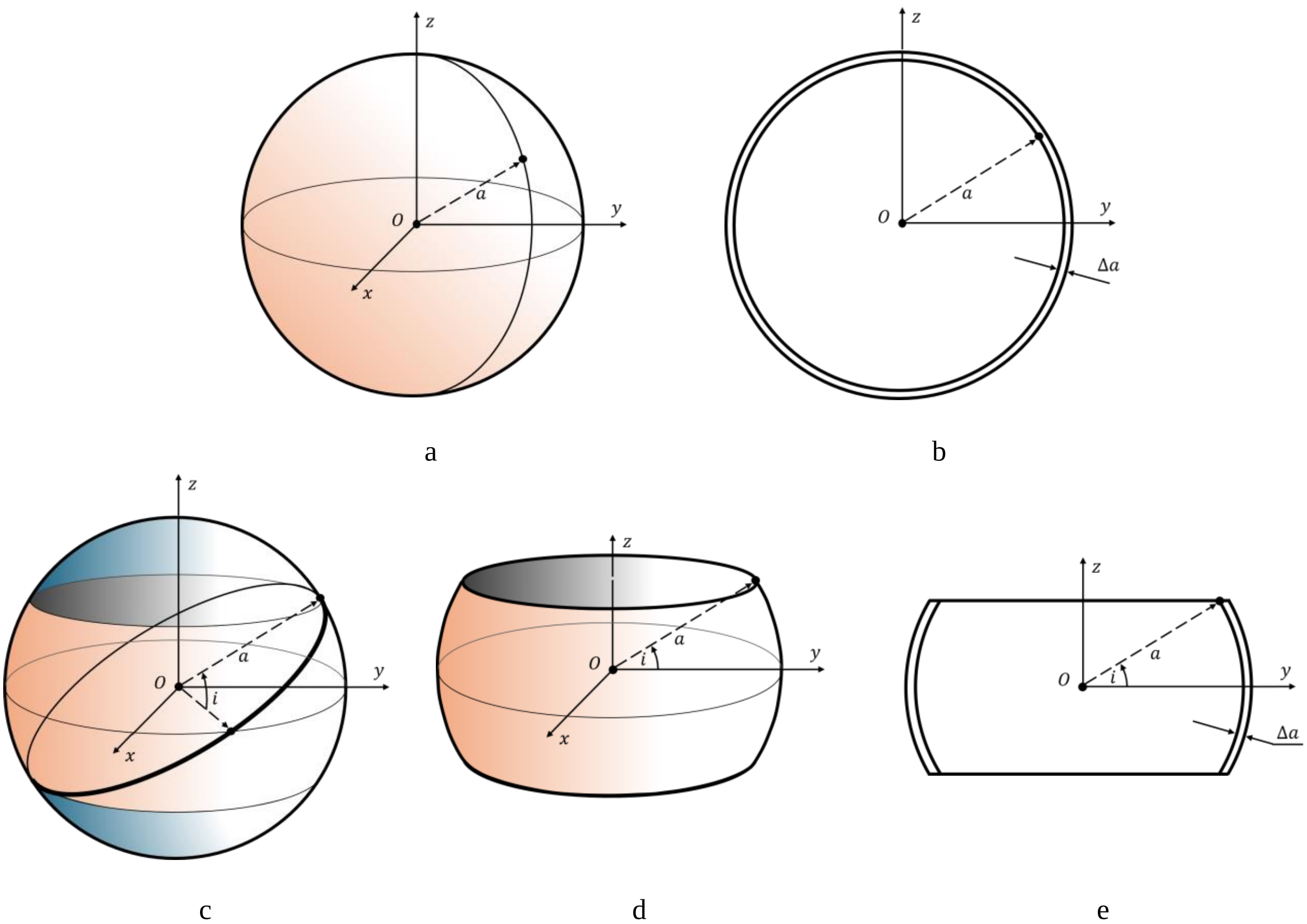

**Fig. 6**. (a) The stability model assumes a satellite is in a circular orbit with semi-major axis, $a$, and spends all its time within the spherical shell (b) with radius, $a$, and thickness, $\Delta a$. However, a satellite in a circular orbit with inclination, $i$, spends all its time within a spherical shell segment, (d), produced by the subtraction from the sphere, (c), of the two spherical caps defined by the orbital inclination, with (e) the corresponding spherical shell segment.

The choice of key stability model parameters also produces results that may be optimistic. The collision risk mitigation manoeuvres are assumed to reduce the breakup rate of satellites to a level that is 1% of the original, unmitigated rate (i.e., $\alpha = 0.01$). This corresponds to a probability of collision threshold for risk mitigation manoeuvres that is likely an order of magnitude or more below the NASA-recommended level of 1E-4[34] and in line with probability of collision thresholds used by some constellation operators[35,36]. Moreover, it is assumed that collision risk mitigation performance is perfect (i.e., $\varepsilon = 1$, see "Methods"). This assumption is consistent with the capability to perform manoeuvres in a timely fashion, conjunction alerts arriving with sufficient warning time, and where these alerts are derived from accurate and reliable sensor or operator data. Delays for the first orbital elements of the fragments of a breakup to enter the public catalogue[37] and the reduction in the accuracy of situational awareness caused by manoeuvring satellites are also ignored[38]. These parameter choices lead to larger critical constellation sizes. Real world risk mitigation performance is imperfect (i.e., $\varepsilon < 1$) and less effective than assumed[39] (i.e., $\alpha > 0.01$), which would tend to reduce the critical sizes of constellations compared with the levels shown.

Additionally, constellation satellites are assumed to perform their post-mission transfer to disposal orbits by adopting a "5-year rule" with a success rate of 99%, which represents compliance rates higher than currently observed in the real world for most satellites[40] but is aligned with levels recommended by the Inter-Agency Space Debris Coordination Committee[41] and NASA[42] for constellations. The model also assumes a "control-to-re-entry" approach consistent with real-world constellation owner/operator practices[31], which enables collision risk mitigation manoeuvres to be performed until the satellites re-enter the atmosphere (or very close to doing so). The results above show consistent benefits of the “control-to-re-entry” approach compared with the 5-year rule. Whilst requiring further study to more fully quantify and understand the benefits and potential drawbacks for constellations, these early results suggest a need to shift away from the traditional post-mission disposal and passivation favoured in international guidelines. Nevertheless, even accounting for these benefits some of the investigated constellations exceeded their critical sizes for runaway and/or unstable environments, indicating a growing breakup rate and fragment generation.

## 4 Methods

The data for constellations and satellites in Tab. 1 and Fig. 1 are from the European Space Agency’s Database and Information System Characterising Objects in Space (DISCOS)[32], FCC filings[9-17], and from data curated by Jonathan McDowell[2]. Where available, the mass and average cross-sectional area from DISCOS are used. There are no satellites deployed for the SpaceX Starmind and Starlink Generation 3 constellations, so the mass and area values from DISCOS for a deployed Starlink satellite (69031) are used. Similarly, Blue Origin TeraWave is yet to be deployed, so the mass and area values from DISCOS for a deployed Kuiper satellite (67158) are used. The numbers of satellites identified in the FCC filings for Starmind[9] and Amazon Leo D2D System[15] are greater than the stated sizes of these constellations of 1,000,000 and 5,105, respectively, so the number of satellites in each orbital shell is scaled to achieve the required numbers. The area of each element in Fig. 1 represents the total number of satellites in the constellation.

Figs. 2-5 are produced using the new stability model parameterised using the values in Tab. 1 for each constellation, or with modified values as described in “Results”. In the original stability model by Kessler and Anz-Meador, the critical number of intact objects leading to a runaway environment is[5],

$$ {}_{R}N_i = \frac{4\pi a^3 V_o \rho_a C_D \left(A_f/m_f\right)}{N_0 W V \sigma_f}, \tag{1} $$

where $a$ is the semi-major axis, $V_o$ the orbital speed assuming a circular orbit, $\rho_a$ the atmospheric density, $C_D$ the drag coefficient, $N_0$ the number of fragments with sufficient mass to catastrophically breakup an intact satellite from the breakup of an intact satellite, $\left(A_f/m_f\right)$ the average area-to-mass ratio of those fragments, $W$ a parameter accounting for the eccentricity of fragment orbits, $V$ the impact speed that converts spatial density to flux, and $\sigma_f$ the combined cross-sectional area between an intact satellite and a fragment. Subsequently, the critical number of intact objects corresponding to an unstable environment is,

$$ {}_{U}N_i = \frac{{}_{R}N_i \sigma_f}{\sigma_f + k\sigma_i}, \tag{2} $$

where $\sigma_i$ is the collision cross-section of two intact objects and $k$ is a factor applied to the intact population.

Satellites that are orbit raising toward a constellation, orbit lowering after retiring from the constellation, or decaying due to atmospheric drag, produce equilibrium spatial densities of intact satellites at altitudes below the constellation. The breakup of these satellites also produces equilibrium fragment spatial densities below the constellation. All these spatial densities are functions of the constellation size, $N_C$. Hence, Eq. 1 is modified to include a function accounting for these terms and collision risk mitigation, thereby giving the critical constellation size,

$$ {}_{R}N_C = {}_{R}N_i \phi_C. \tag{3} $$

The added function is,

$$\Phi_C = \frac{1}{\Delta a\left[(\varepsilon\alpha + 1 - \varepsilon)\left(\frac{1+\eta}{t_L\left(\frac{da}{dt}\right)_R} + \frac{q\beta}{t_L\left(\frac{da}{dt}\right)_L} + \frac{1}{\Delta a}\right) + \frac{(1-q)\gamma}{t_L\left(\frac{da}{dt}\right)_{PMD}} + \frac{q(1-\beta) + (1-q)(1-\gamma) + \eta}{4\pi a^2 t_L\left(\frac{da}{dt}\right)_D}\right]}. \tag{4}$$

where $\eta$ is the proportion of satellites that fail, $t_L$ is the satellite lifetime, $\Delta a$ is the thickness of the spherical shells, $(da/dt)_R$ is the rate of change of the semi-major axis during orbit raising, $(da/dt)_L$ is the rate of change of the semi-major axis during orbit lowering, and $(da/dt)_D$ is the rate of change of the semi-major axis due to atmospheric drag. Operational satellites are assumed to manoeuvre with success rate, $\varepsilon$, to reduce the probability of collision to a proportion, $\alpha$, of the unmitigated value. Eq. 4 includes a Boolean parameter, $q \in \{0,1\}$, used to select a post-mission disposal option. Satellites reaching their end-of-life can perform post-mission disposal via a "control-to-re-entry" (C2R) approach ($q = 1$) with success rate, $\beta$, which also maintains their ability to manoeuvre, or via transfer to a disposal orbit with success rate, $\gamma$, to meet a required disposal timeline, $t_{PMD}$, with subsequent passivation that removes the ability to manoeuvre to mitigate collision risks ($q = 0$). The proportion of satellites failing to complete their post-mission disposal successfully $(1-\gamma)$ are assumed to be abandoned and will decay only under the influence of atmospheric drag. Hence, the total proportion of the satellites in the constellation that fail, is $(1-\gamma)+\eta$. See "Supplementary Materials" for a full derivation of the stability model.

Eq. 3 is used to estimate the critical constellation size for the 14 constellations in Tab. 1 at all altitudes between 200 km and 2000 km with altitude bin size, $\Delta a = 10$ km. Additionally, the stability model is used to estimate the critical constellation size for the Blue Origin Project Sunrise constellation at 1,800 km altitude. Atmospheric densities for each bin were from the CIRA-72 atmospheric model[43] under the assumption of an average solar activity of 130, representing an exospheric temperature of 980 K after accounting for geomagnetic activity. Following Lewis and Kessler[6], $N_0 = 112$ fragments per breakup, $W = 1.1$, $V = 7{,}500$ ms$^{-1}$, and $C_D = 2.2$. The parameter $k$ is assumed to take the value 2.

For every constellation and every satellite, a "control-to-re-entry" post-mission disposal approach ($q = 1$) is adopted with success rate, $\beta = 1$ at their end of life ($t_L = 5$ years). Operational satellites perform orbit raising at rate $(da/dt)_R = 0.05$ ms$^{-1}$ and orbit lowering to an altitude of 125 km at rate, $(da/dt)_L = 0.05$ ms$^{-1}$. Operational satellites are assumed to perform collision risk mitigation manoeuvres with success rate, $\varepsilon = 1$, which reduce the collision rate to a factor $\alpha = 0.01$ of the original rate. The proportion of satellites assumed to fail is, $\eta = 0.01$. The failure rate is in line with the rate recorded by Jonathan McDowell for all Starlink satellites launched ($\eta = 0.013$)[18] but is higher than the rate recorded for OneWeb satellites ($\eta = 0.003$)[44]. Note that relatively few Starlink failures have affected the de-orbiting or manoeuvring capability of the satellites.

A second scenario is used in which every constellation adopts a 5-year (5YR) post-mission disposal approach, as required by the FCC [22,29]. Constellation satellites are assumed to transfer to a disposal orbit ($q = 0$) with residual orbital lifetime, $t_{PMD} = 5$ years, and success rate, $\gamma = 1$ at their end of life ($t_L = 5$ years). With the proportion of satellites assumed to fail, $\eta = 0.01$, the total post-mission disposal success rate is 99%, in keeping with the IADC[41] and NASA[42] recommendations. Other parameters are as used for the "control-to-re-entry" scenario. The assumptions for both post-mission disposal scenarios reflect internationally recognised good practices and real-world operator practices surpassing these.

## 5 Acknowledgements

The author thanks Jonathan McDowell for help with access to constellation data and acknowledges use of the European Space Agency's Database and Information System Characterising Objects in Space and the Federal Communications International Communication (FCC) Filing System, and the efforts of the teams and individuals involved in their maintenance and operation. The author also thanks Sarah Lewis for feedback on an earlier version of the manuscript. Finally, the author acknowledges the invaluable mentorship and guidance of Donald Kessler and Phillip Anz-Meador.

## 6 Competing interests

The author declares no competing interests.

## 7 Supplementary Materials

### Stability Model Derivation

Assume the population of operational constellation satellites, $N_C$, is in equilibrium through regular replenishment and disposal of individual satellites. If there are always $N_C$ operational satellites in the constellation, each with operational lifetime, $t_L$, and the proportion of satellites that fail is $\eta$, then the average launch rate needed to maintain the constellation in equilibrium is,

$$\frac{dL}{dt} = \frac{N_C(1+\eta)}{t_L}. \tag{1}$$

Satellites reach their end-of-life without failing at rate,

$$\frac{dE}{dt} = \frac{N_C}{t_L}, \tag{2}$$

and satellites fail at rate,

$$\frac{dF}{dt} = \frac{\eta N_C}{t_L}. \tag{3}$$

The recurring orbit raising, orbit lowering, and decaying of constellation satellites generate populations below the constellation with altitude- and time-dependent spatial densities, $S_1$. Integrating these spatial densities over the time taken by the satellites to traverse through different altitudes produces average, or equilibrium, spatial densities. For example, the equilibrium spatial density of satellites that are orbit raising toward the constellation is,

$$S_R(a) = \frac{dL}{dt}\int S_{1,R}(a,t)\,dt = \frac{N_C(1+\eta)}{t_L}\int S_{1,RAISE}(a,t)\,dt. \tag{4}$$

The spatial density $S_1$ for a single satellite is

$$S_1 = \frac{\Delta t}{T\Delta U}, \tag{5}$$

where $\Delta t$ is the time spent in a volume element $\Delta U$ over time $T$. Time $T$ is assumed to be long enough such that all regions of space accessible to the satellite have been traversed. Assuming a spherical shell of radius $R$ and thickness $\Delta R$, with $\Delta R \ll R$, between the perigee and apogee of a satellite's orbit, the volume of the shell is the surface area of the sphere with radius $R$ times the thickness of the shell,

$$\Delta U = 4\pi R^2 \Delta R. \tag{6}$$

Assuming the satellite is in a circular orbit with semi-major axis $a = R$, $\Delta t/T = 1$, and the shell thickness is $\Delta a = \Delta R$, the spatial density averaged over all latitudes is,

$$S_1 = 1/4\pi a^2 \Delta a. \tag{7}$$

Hence, for a single satellite the integral of its spatial density is,

$$\int S_1(a,t)\,dt = \int \frac{1}{4\pi a^2 \Delta a}\,dt = \frac{1}{4\pi a^2 (da/dt)}. \tag{8}$$

The rate of change of the semi-major axis, $da/dt$, depends upon whether the satellite is using any onboard propulsion for orbit raising or lowering, or whether the satellite is only subject to atmospheric drag. Examination of Two-Line Element (TLE) data for the satellite Starlink-1780 (46684), which used onboard propulsion for orbit raising and lowering, suggests the following rates of change for orbit raising and lowering:

$$(da/dt)_R = 0.08\ \mathrm{ms}^{-1},\ (da/dt)_L = 0.04\ \mathrm{ms}^{-1}. \tag{9}$$

For a satellite subject only to atmospheric drag, the rate of change of the semi-major axis is a function of the atmospheric density,

$$(da/dt)_D = a\rho_a C_D V_o (A_i/m_i), \tag{10}$$

where $\rho_a$ is the atmospheric density at altitude $h_1$, $C_D$ is the drag coefficient (assumed to be 2.2), $V_o$ is the orbital velocity at altitude $h = a - r_E$, and $(A_i/m_i)$ is the area-to-mass ratio of the intact satellite.

Combining Eqs. 4 and 8 gives the average, or equilibrium, density of operational satellites that are orbit raising:

$$S_R = \frac{N_C(1+\eta)}{4\pi a^2 t_L (da/dt)_R}. \quad (11)$$

Similarly, the equilibrium density of operational satellites that are using onboard propulsion for orbit lowering is,

$$S_L = \frac{N_C}{4\pi a^2 t_L (da/dt)_L}, \quad (12)$$

and the equilibrium density of failed satellites with orbits decaying due to atmospheric drag is,

$$S_D = \frac{\eta N_C}{4\pi a^2 t_L (da/dt)_D} = \frac{\eta N_C}{4\pi a^3 t_L \rho_a C_D V_o (A_i/m_i)}. \quad (13)$$

Eq. 13 shows an important, inverse proportionality with the atmospheric density. At high altitudes where the atmospheric density is low, the average spatial density will be correspondingly high. Further, at most altitudes above about 500 km, the rate of change of semi-major axis due to atmospheric drag will be orders of magnitude lower than the rates of change of semi-major axis produced by propulsion systems. This means that the proportion of satellites that fail will be the key determinant of the equilibrium spatial density.

Finally, the equilibrium density of operational satellites in the constellation providing user services is,

$$S_C = \frac{N_C}{\Delta U} = \frac{N_C}{4\pi a^2 \Delta a}. \quad (14)$$

Given the equilibrium conditions for the constellation, the collisional breakup rate, $dB/dt$, of constellation satellites is constant and producing fragments that decay due to atmospheric drag. Each collision is assumed to produce a time-dependent spatial density of fragments, $S_{1,F}$, below the constellation. The average spatial density of fragments is then,

$$S_B(a) = \frac{dB}{dt} \int S_{1,F}(a,t)\, dt. \quad (15)$$

As before, Eq. 15 is integrated over the time it takes for the fragments to pass through each altitude. The integral in Eq. 15, assuming $N_0$ fragments are produced by a single breakup, is

$$\int S_{1,F}(a,t)\, dt = \frac{N_0 W}{4\pi a^2 (da/dt)} = \frac{N_0 W}{4\pi a^3 \rho_a C_D V_o \left(A_f/m_f\right)}, \quad (16)$$

where $\left(A_f/m_f\right)$ is the average area-to-mass ratio of fragments with sufficient mass to catastrophically breakup an intact constellation satellite. The value of $W$ is 1 for circular orbits and can be as high as 50 for some highly elliptical orbits.

For a collision to be catastrophic at 10 km/sec, the ratio of target mass, $m_i$, to fragment mass, $m_f$, should be less than 1250. Assuming the fragments are spherical with diameter, $d_f > 1$ cm, mass, $m_f = m_i/1250$, and with density,

$$\rho_f = 92.937 {d_f}^{-0.74}, \tag{17}$$

then the fragment area is,

$$A_f = 0.01775 \left(\frac{m_i}{1250}\right)^{-0.74}. \tag{18}$$

Assume constellation satellites can implement collision risk mitigation manoeuvres with a success rate, $[\varepsilon_i] = [\varepsilon_C, \varepsilon_R, \varepsilon_L, \varepsilon_D]$, which reduce the collision rate by $\alpha \in [0,1]$, and that through constellation design most satellite-on-satellite encounters are passively deconflicted such that only a proportion, $[\psi_i] = [\psi_C, \psi_R, \psi_L, \psi_D]$, can result in a breakup. Noting that if two satellites collide, two fragment clouds are produced, the breakup rate is,

$$\frac{dB}{dt} = \alpha[\varepsilon_i][\psi_i][S_i][S_i]^{\mathrm{T}}\sigma_i V\Delta U + [1-\varepsilon_i][\psi_i][S_i][S_i]^{\mathrm{T}}\sigma_i V\Delta U + \alpha[\varepsilon_i][S_i]S_f\sigma_f V\Delta U + [1-\varepsilon_i][S_i]S_f\sigma_f V\Delta U, \tag{19}$$

where the array $[S_i] = [S_C, S_R, S_L, S_D]$ is the spatial density of constellation satellites in different mission phases, and $S_f$ is the fragment spatial density, in the spherical shell segment with volume $\Delta U$. $V$ is the average relative velocity that transforms spatial density into flux and takes an assumed, conservative value of 7,500 ms$^{-1}$, $\sigma_i$ is the combined collision cross-sectional area between two intact satellites,

$$\sigma_i = \left(A_i^{1/2} + A_i^{1/2}\right)^2 = 4A_i, \tag{20}$$

and $\sigma_f$ is the combined collision cross-sectional area between an intact satellite and a fragment,

$$\sigma_f = \left(A_i^{1/2} + A_f^{1/2}\right)^2. \tag{21}$$

Finally, assume that the environment is in equilibrium and the altitude of the volume above $h$ is similar to $h$, leading to the approximation, $S_B = S_f$. Hence, combining Eqs. 15 and 19 gives the equilibrium spatial density of fragments,

$$S_B = \frac{\left(\alpha[\varepsilon_i][\psi_i][S_i][S_i]^{\mathrm{T}}\sigma_i V\Delta U + [1-\varepsilon_i][\psi_i][S_i][S_i]^{\mathrm{T}}\sigma_i V\Delta U\right)\int S_{1,F}(a,t)\,dt}{1-\left(\alpha[\varepsilon_i][S_i]\sigma_f V\Delta U + [1-\varepsilon_i][S_i]\sigma_f V\Delta U\right)\int S_{1,F}(a,t)\,dt}. \tag{22}$$

As the denominator approaches zero, $S_B$ tends to infinity. Therefore, combining Eq. 22 and Eq. 16 gives the condition for a runaway fragment population, caused solely by the deployment of the constellation,

$$\frac{N_0 W}{4\pi a^3 \rho_a C_D V_o\left(A_f/m_f\right)}\left(\alpha[\varepsilon_i][S_i]\sigma_f V\Delta U + [1-\varepsilon_i][S_i]\sigma_f V\Delta U\right) = 1. \tag{22}$$

Substituting the spatial densities of the orbit raising, orbit lowering, user-serving, and decaying satellites from Eqs. 11-14, assuming $\varepsilon_C = \varepsilon_R = \varepsilon_L = \varepsilon$ and $\varepsilon_D = 0$, and rearranging, gives the critical constellation size, ${}_R N_c$, for a runaway environment:

$${}_R N_c = \frac{4\pi a^3 V_o \rho_a C_D \left(A_f/m_f\right)}{N_0 W V \sigma_f}\Phi_C \tag{23}$$

where,

$$\Phi_C = \frac{1}{\Delta a\left[(\varepsilon\alpha + 1 - \varepsilon)\left(\frac{1+\eta}{t_L\left(\frac{da}{dt}\right)_R} + \frac{1}{t_L\left(\frac{da}{dt}\right)_L} + \frac{1}{\Delta a}\right) + \frac{\eta}{4\pi a^2 t_L\left(\frac{da}{dt}\right)_D}\right]}. \tag{24}$$

We can further modify Eq. 24 to account for more traditional post-mission disposal approaches that use transfers to disposal orbits to achieve a specific residual orbital lifetime (e.g., 5 years), before the satellites are passivated. We can also add success rates for these post-mission disposal approaches.

Firstly, we introduce a Boolean parameter, $q \in \{0,1\}$, where $q = 1$ for constellation satellites that use propulsion to lower the orbit until it re-enters (a process we refer to as "control-to-re-entry"), and $q = 0$ for constellation satellites that transfer to a disposal orbit with residual orbital lifetime, $t_{PMD}$, and are then passivated. For this latter case, the satellites no longer have manoeuvring capability after being passivated and decay from orbit due to the influence of atmospheric drag only. If the proportion of constellation satellites that complete the control-to-re-entry process is $\beta$ and the proportion of satellites that complete a transfer to a disposal orbit is $\gamma$, then Eq. 24 becomes,

$$\Phi = \frac{1}{\Delta a\left[(\varepsilon\alpha + 1 - \varepsilon)\left(\frac{1+\eta}{t_L\left(\frac{da}{dt}\right)_R} + \frac{q\beta}{t_L\left(\frac{da}{dt}\right)_L} + \frac{1}{\Delta a}\right) + \frac{(1-q)\gamma}{t_L\left(\frac{da}{dt}\right)_{PMD}} + \frac{q(1-\beta) + (1-q)(1-\gamma) + \eta}{4\pi a^2 t_L\left(\frac{da}{dt}\right)_D}\right]}, \tag{25}$$

where

$$\left(\frac{da}{dt}\right)_{PMD} = \max\left\{\left(\frac{da}{dt}\right)_D, \frac{\Delta H}{t_{PMD}}\right\}, \tag{26}$$

with $\Delta H$ the difference in altitude between $h$ and 125 km, which is the altitude assumed for re-entry. Eq. 26 captures the cases in which the atmospheric density is sufficiently large to cause re-entry in times less than $t_{PMD}$. In general, the value of $(da/dt)_{PMD}$ is orders of magnitude less than $(da/dt)_L$, except at very low altitudes, meaning that disposal via control-to-re-entry will lead to higher thresholds for a runaway environment. Finally, we assume in Eq. 25 that satellites failing to complete their post-mission disposal successfully are abandoned and will decay only under the influence of atmospheric drag at rate $(da/dt)_D$ calculated from Eq. 10. As such, the total proportion of the satellites in the constellation that fail, is $(1-\gamma) + \eta$.